\documentclass[a4paper,11pt]{article}
\pdfoutput=1 

\usepackage{jinstpub} 
\usepackage{changepage}
\usepackage{xcolor}
\usepackage{graphicx}
\usepackage{xspace}
\usepackage[utf8]{inputenc}
\usepackage[english]{babel}
\usepackage{caption}
\usepackage{subcaption}
\usepackage{float}
\usepackage{blindtext}
\usepackage[colorlinks = true,
            linkcolor = blue,
            urlcolor  = blue,
            citecolor = blue,
            anchorcolor = blue]{hyperref}
\usepackage[capitalise,noabbrev]{cleveref}
\usepackage{chngcntr}
\usepackage{tabularray}
\usepackage{lineno}

\counterwithin{figure}{section}

\addto\captionsenglish{}

\title{A Sub-Picosecond Digital Clock Monitoring System}

\author[a]{Rohith Saradhy}
\author[a]{Erich Frahm}
\author[b]{Eduardo B.S. Mendes}
\author[a]{Roger Rusack}

\affiliation[a]{The University of Minnesota,\\Minneapolis, Minnesota, USA}
\affiliation[b]{CERN, Geneva, Switzerland}

\emailAdd{rusack@umn.edu,rsaradhy@umn.edu}

\begin{document}
\abstract{We describe a low-cost system designed to monitor wander in digital clocks with a precision of $\le$ 1 ps. With this system we have shown that it is possible to track phase variations at the sub-picosecond level by adding noise to a reference clock. As in many cases where a clock is part of a complex distribution network small changes in temperature and other effects can lead to small changes in the clock's phase. As a further demonstration of the system, we have used it to measure the phase changes induced in optical signals in fibers.}  

\maketitle


\section{Introduction}

Systems that can distribute precision reference clocks stable to a level of a few picoseconds is a common theme in many current and proposed experiments in high energy physics. 
At the CERN High-Luminosity LHC (HL-LHC) the average number of interactions in each bunch crossing, will be 140 or higher. To help contend with this level of pile up, both the ATLAS and CMS Collaborations are planning to install specialized detectors capable of measuring the arrival time of a particle to 30~ps or less. 

For these detectors and others that exploit precision timing, both distributing and monitoring the stability of the reference clock will be an essential task. 
In this paper we describe a clock-monitoring system that we have developed to monitor drifts in the reference clock with sub-picosecond precision. 


At the sub-picosecond level, the use of currently available frequency counters or time interval counters are precluded. One method for comparing two clocks employs a scheme known as the Dual Time-Difference Measurement circuit, where the clock signals are heterodyned and the beat clock used to extract the time difference \cite{allan1}. A digital version of this circuit, the digital dual mixer time difference (DDMTD) circuit, was first proposed by Moriera and Darwazeh \cite{Moreira2011DigitalFT}. The circuit has recently been implemented 
in an FPGA \cite{Mendes_1} and with phase stabilization has achieved a level of $\approx$ 2~ps \cite{Mendes_2}. In this paper we report how we have used this approach with modern discrete RF components to design a system to measure variations in a clock's phase to less than 100~fs.

In this report we describe the basic principle of operation of a DDMTD Circuit and report on results obtained from the laboratory measurements. We show that it is feasible to detect sub-picosecond phase variations with such a system.


\section{Principle of Operation \label{theory}}

\begin{figure}
    \includegraphics[width=\textwidth,keepaspectratio]{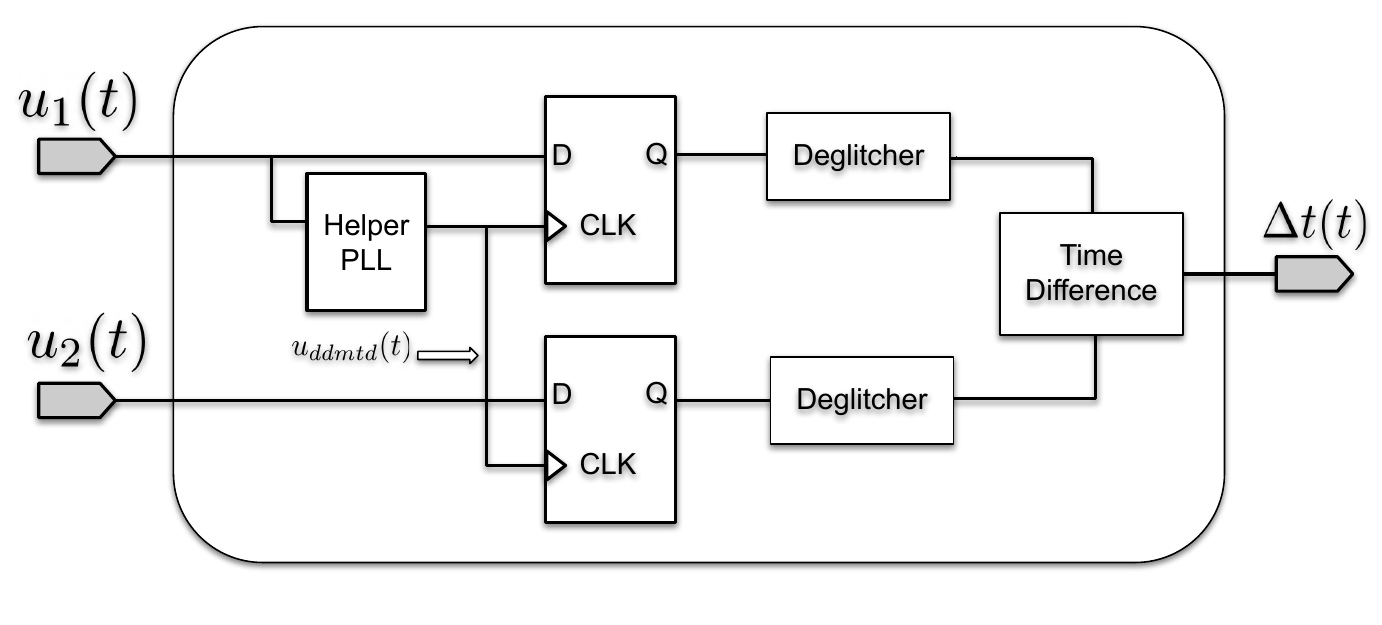}
    
    \caption{\label{fig:schematic} Schematic of DDMTD from~\cite{Moreira2011DigitalFT}.}
\end{figure}

The digital dual mixer time difference (DDMTD) is a digital circuit composed of a phase-locked-loop (PLL) 
and two flip-flops (FF) that can be used to compare the time interval error (TIE) of two clocks with 
high precision.  
The circuit is shown schematically in \cref{fig:schematic}.  
There are two input clocks, \textit{\large $u_{1}$} and \textit{\large $u_{2}$}, with frequencies \textit{\large $\nu_{1}$} and \textit{\large $\nu_{2}$} respectively; Where \textit{\large $u_{1}$} is the reference clock and \textit{\large $u_{2}$} is the test clock that is to be compared with the reference clock. 

To track variations in phase between the two clocks a 'helper' PLL creates a new clock \textit{\large $u_{ddmtd}$} that is phase-locked to \textit{\large $u_{1}$} and has a frequency (\textit{\large $\nu_{ddmtd}$}) that is slightly offset from \textit{\large $\nu_{1}$}. The choice of the offset frequency is chosen using  \cref{nu_ddmtd}:

\label{working-eqns-section}

\begin{equation}
    \nu_{ddmtd} = \dfrac{N}{N+1} . \nu_{1} 
    \label{nu_ddmtd}
\end{equation}

\noindent Here N is an integer that determines the number of input clock cycles required for a full phase cycle of the heterodyned signal.

The input clocks, \textit{\large $u_{1}$} and \textit{\large $u_{2}$}, are compared by sampling them with two D-type Flip-Flops clocked on the positive edge of \textit{\large $u_{ddmtd}$}. Whenever the phase between \textit{\large $u_{ddmtd}$} and the input clock changes by some integer multiple of $\pi$, the output of the flip-flops will change state, tracking the integrated time difference over a period of N/2 cycles of \textit{\large $u_{ddmtd}$} clock. The resulting output of the flip-flops will form a clock with beat frequency given by \cref{beat_freq}. The phases of these beat clocks will directly depend on the relative phases of clocks \textit{\large $u_{1}$} and \textit{\large $u_{2}$}.





\begin{equation}
    \label{beat_freq}
    \nu_{beat} = \dfrac{1}{N} . \nu_{ddmtd}  = \dfrac{1}{N+1} . \nu_{1} 
\end{equation}

\noindent The integrated time difference between two clock edges can then be measured after N/2 input clock cycles with a minimum precision given by \cref{T_Res}. 
\begin{equation}
    \label{T_Res}
    \Delta t _{min} = \dfrac{1}{\nu_{ddmtd}}. \dfrac{\nu_{beat}}{\nu_{1}} = \dfrac{t_{ddmtd}}{N+1} = \dfrac{t_{1}}{N} 
\end{equation}

Here, the $\nu_{1} / \nu_{ddmtd} = N+1$ is the effective temporal gain ($T_g$) of the DDMTD. By measuring the difference between the two beat clocks, the phase difference between \textit{\large $u_{1}$} and \textit{\large $u_{2}$} can be measured to a precision given by  \cref{Time_Diff}  \cite{Moreira2011DigitalFT}.

\begin{equation}
    \label{Time_Diff}
    \Delta t = \Delta t_{beat}. \dfrac{\nu_{beat}}{\nu_{1}} =  \dfrac{\Delta t_{beat}}{N+1}
\end{equation}

Here, $\Delta t_{beat}$ is the time difference between the transitions of the two beat clocks and $\Delta t$ is the time difference between the input clocks. Since $T_g$ is directly proportional to N, the limit of the temporal gain depends on how close of a frequency the helper PLL can accurately generate with respect to the input clock frequency.

For the results reported in this paper, we have used digital input clocks with a frequency of 160~MHz and N=100k or10k, with a theoretical precision of 62.5~fs and 625~fs respectively.



As the Time Interval Error (TIE) is determined from both the positive and negative edges of the beat clocks, the highest frequency that can be measured is given by \cref{freq_max}, with $\nu_{max}$ approximately 1.6~kHz and 16~kHz for N=100k, and 10k respectively. 

\begin{equation}
    \label{freq_max}
    \nu_{max} < \dfrac{2_{pos-negedge} \times \nu_{beat}}{2_{Nyquist Limit}} =  \dfrac{\nu_{1}}{N+1} 
\end{equation}

When the \textit{\large $u_{ddmtd}$} has a transition edge close to the transition edge of the input clocks, 
the set-up and hold times will be violated and the output of the flip-flop becomes unstable.
This instability can last for several clock cycles of the input clock. 
To estimate the exact time of the transition, the average of the time of the first and last transitions can be used. The length of this instability is determined by the setup and hold times of the flip-flops used. Standard CMOS devices have setup and hold times of $\sim 800$~ps, while for the silicon-germanium NB7V52M flip-flops used in the circuit discussed below, they are $\sim 15$~ps. Thus the exact time of the transition can be more accurately estimated.

\section{System Design}

\begin{figure}[htb]
   \centering
   \includegraphics[width=0.7\linewidth]{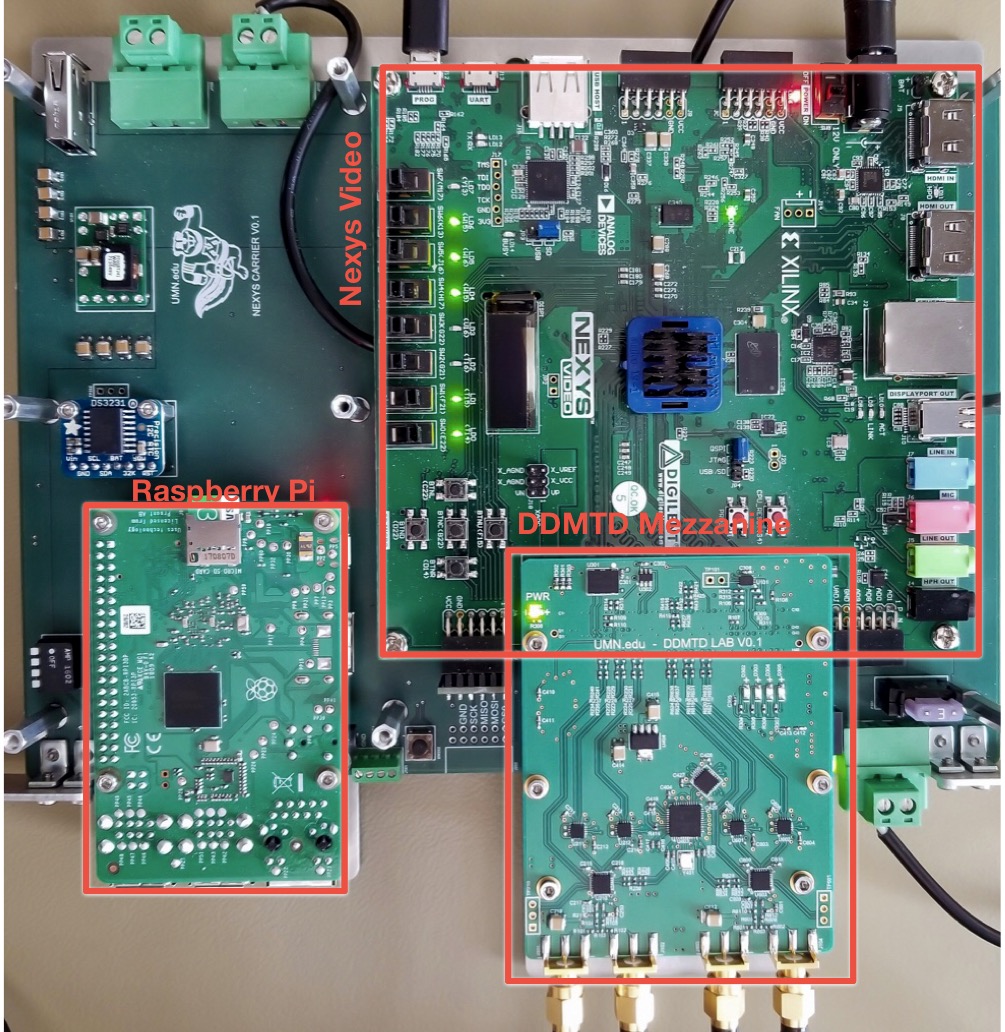}
   \caption{DDMTD Nexys Board.} \label{DDMTD_pic}
\end{figure}

The clock-monitoring system that we developed to investigate the precision that can be achieved is shown in \cref{DDMTD_pic}. It is based on a 5U motherboard, which supports a Nexys Video board and a Raspberry Pi, with the DDMTD circuit mounted in a mezzanine board connected to the Nexys Video board via an FMC connector. The Raspberry Pi manages the configuration of the on-board electronics and the data acquisition from the Artix-7 FPGA on the Nexys Video board. The beat clocks from the DDMTD mezzanine card are sampled by the FPGA and stored in memory until the Raspberry Pi starts data acquisition.


\begin{figure}[h!]
   \centering
   \includegraphics[width=0.7\linewidth]{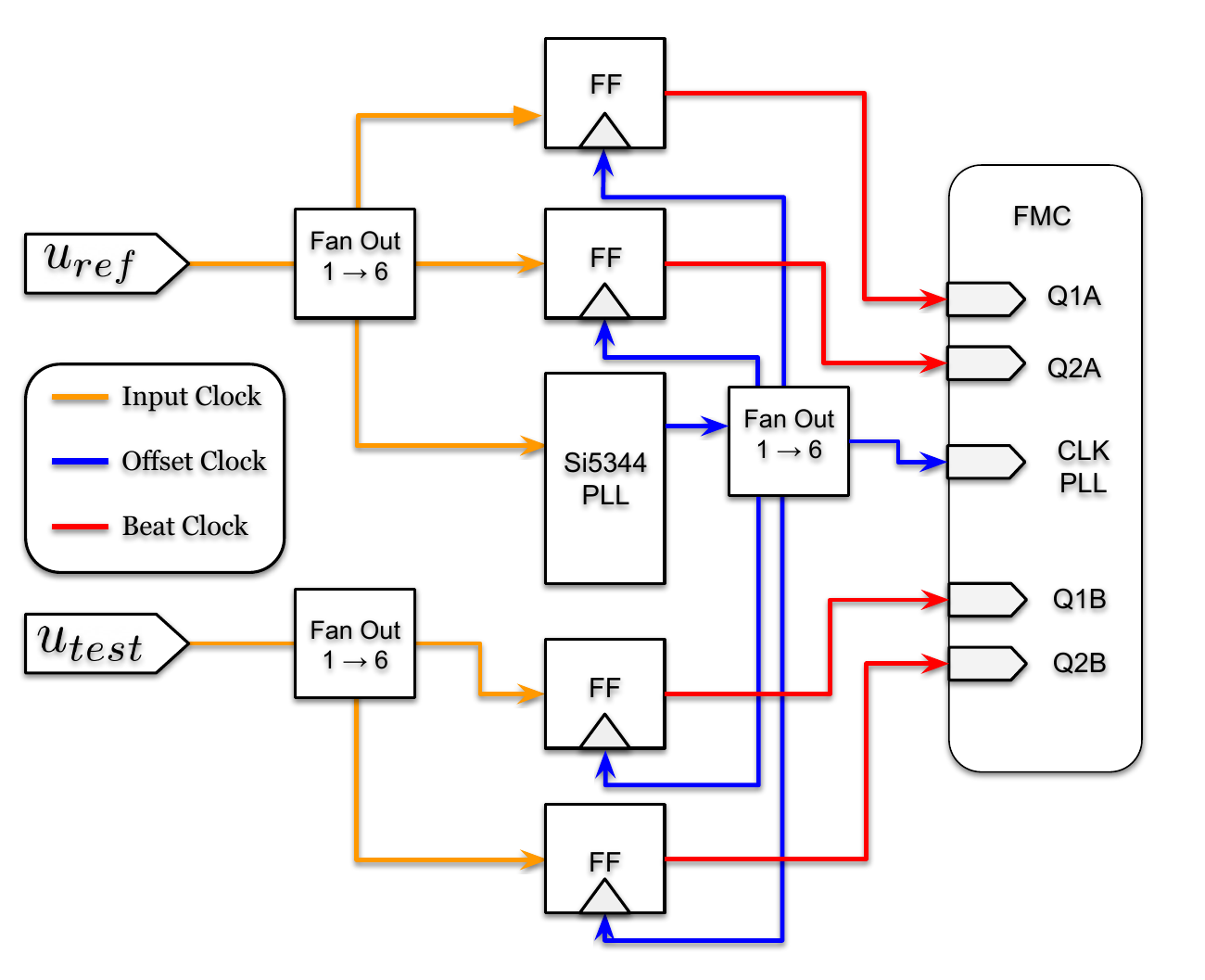}
   \caption{Schematic of the DDMTD Mezzanine. The clocks $u_{ref}$ and $u_{test}$ are fed into flip flops that are clocked by an offset clock generated by a SI5344 Jitter Attenuator / Clock Multiplier. The outputs of the flip-flops are sent via the FMC connector to the Artix-7 FPGA on the Nexys video board.} \label{DDMTD-schema}
\end{figure}

A schematic of the DDMTD mezzanine is shown in \cref{DDMTD-schema}. The circuit is equipped with four high-performance flip-flops (FF),\footnote{NB7V52M, manufactured by ON-Semiconductors} which have a maximum clocking rate of 12~GHz and a Silicon Labs Si5344 Jitter Attenuator / Clock Multiplier IC that has a quoted rms jitter of 90~fs.\footnote{https://www.silabs.com/documents/public/data-sheets/Si5345-44-42-D-DataSheet.pdf} The choice of duplicating the two input flip-flops was to allow the investigation of the properties and stability of the flip-flops. 


The two input clocks are fed into the mezzanine board as differential pairs on SMA connectors. The clocks are fanned out to the four flip-flops and to the Si5344 using the 1-to-6 fan-out chip NB7VQ1006M from ON Semiconductors. These fan-outs are high-performance with a typical RJ (Random Jitter)+DJ (Deterministic Jitter) = (0.2 + 3.0)~ps.\footnote{https://www.onsemi.com/pub/Collateral/NB7VQ1006M-D.PDF} The differential D-type flip-flop NB7V52M has an RMS jitter $< 0.8$ ps.\footnote{https://www.onsemi.com/pub/Collateral/NB7V52M-D.PDF}
%
%
The output of the helper PLL Si5344 is distributed to the clock inputs of all the flip-flops using a second NB7VQ1006M. A copy of this clock (CLK PLL) is transmitted to the Artix-7 FPGA to drive the sampling logic in the FPGA. 

\begin{figure}[h]
   \centering~
   \includegraphics[width=0.9\linewidth]{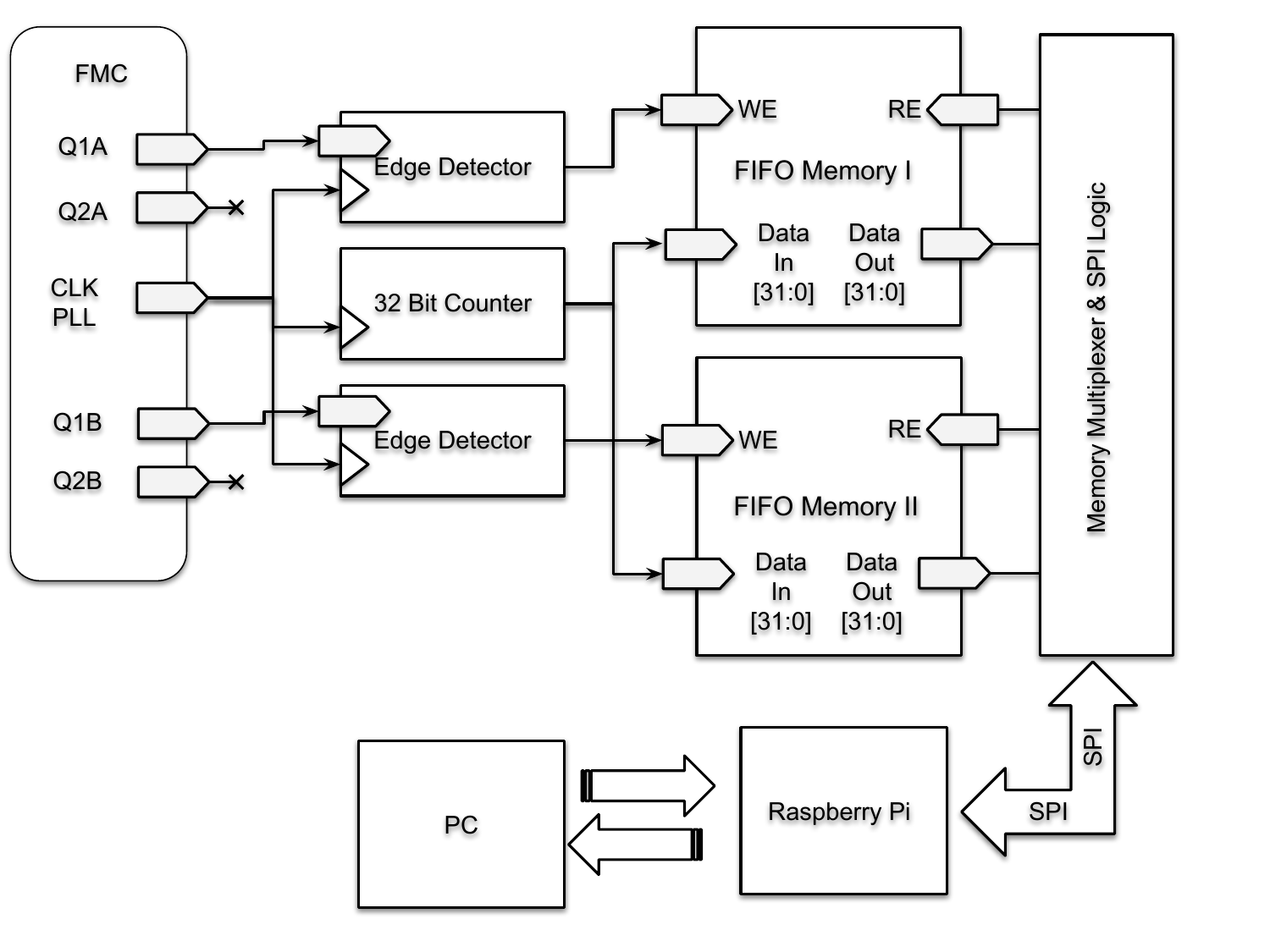}
   \caption{Schematic of the FPGA Logic.} \label{FPGA_Logic}
\end{figure}

The firmware logic implemented on the Artix-7 FPGA is shown in \cref{FPGA_Logic}. Two FIFOs are used to store the value of a 32-bit counter, which is incremented by the CLK PLL. Whenever one of the beat clocks, Q1A or Q1B, changes state the value of the counter is pushed to the FIFO. When the FIFO chain is almost full, the Raspberry Pi pulls the data through the SPI bus and sends it to the PC via Ethernet.

Offline a Fast Fourier Transform (FFT) is used to determine the frequency components of the jitter. The exact time of the phase transition between the two clocks is estimated from the difference between the first and the last edges of the meta-stable region.


\section{Characterizing System Performance by Injecting Noise}

To investigate the performance of the DDMTD and to measure the precision with which TIE can be measured, sinusoidal jitter patterns were injected onto a 160~MHz digital clock. For this two 160 MHz clocks were generated by a pulse pattern generator,\footnote{Keysight 81134A} and jitter was injected onto one of them using a secondary function generator\footnote{AIM TTI TG5011A} connected to the Delay Control Input of the pulse generator. While the nominal amplitude of the jitter generated by the function generator for a 1V input signal was 25~ps, all our measurements were consistent with $\sim$12\% calibration error, such that a 1V signal produced a $\approx$ 28~ps jitter.

\begin{figure}
    \centering
    \includegraphics[width=0.95\linewidth]{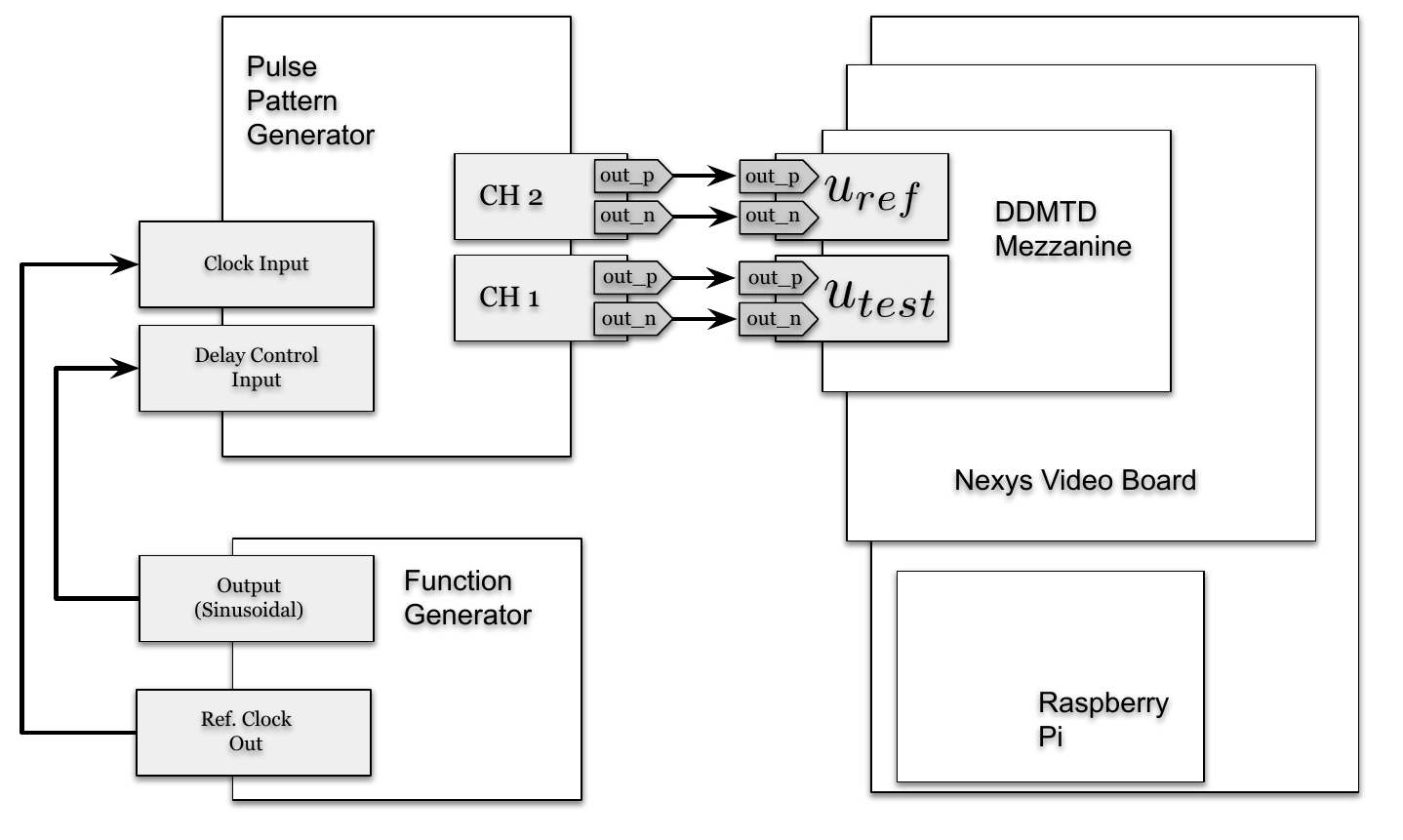}
    \caption{Clock and jitter injection system used in the characterization of the DDMTD.}
    \label{fig:Noise_injector}
\end{figure}

The injected jitter pattern was compared to the jitter pattern recovered using the DDMTD-Nexys Board.
The schematic of the configuration used for the tests is shown in \cref{fig:Noise_injector}. With this set up we were able to investigate sinusoidal jitter injection patterns with amplitudes between 0.25~ps and 25~ps and frequencies up to 6~kHz.  

\begin{figure}
    \begin{minipage}[t]{0.48\linewidth}
        \centering
        \includegraphics[width=0.95\linewidth]{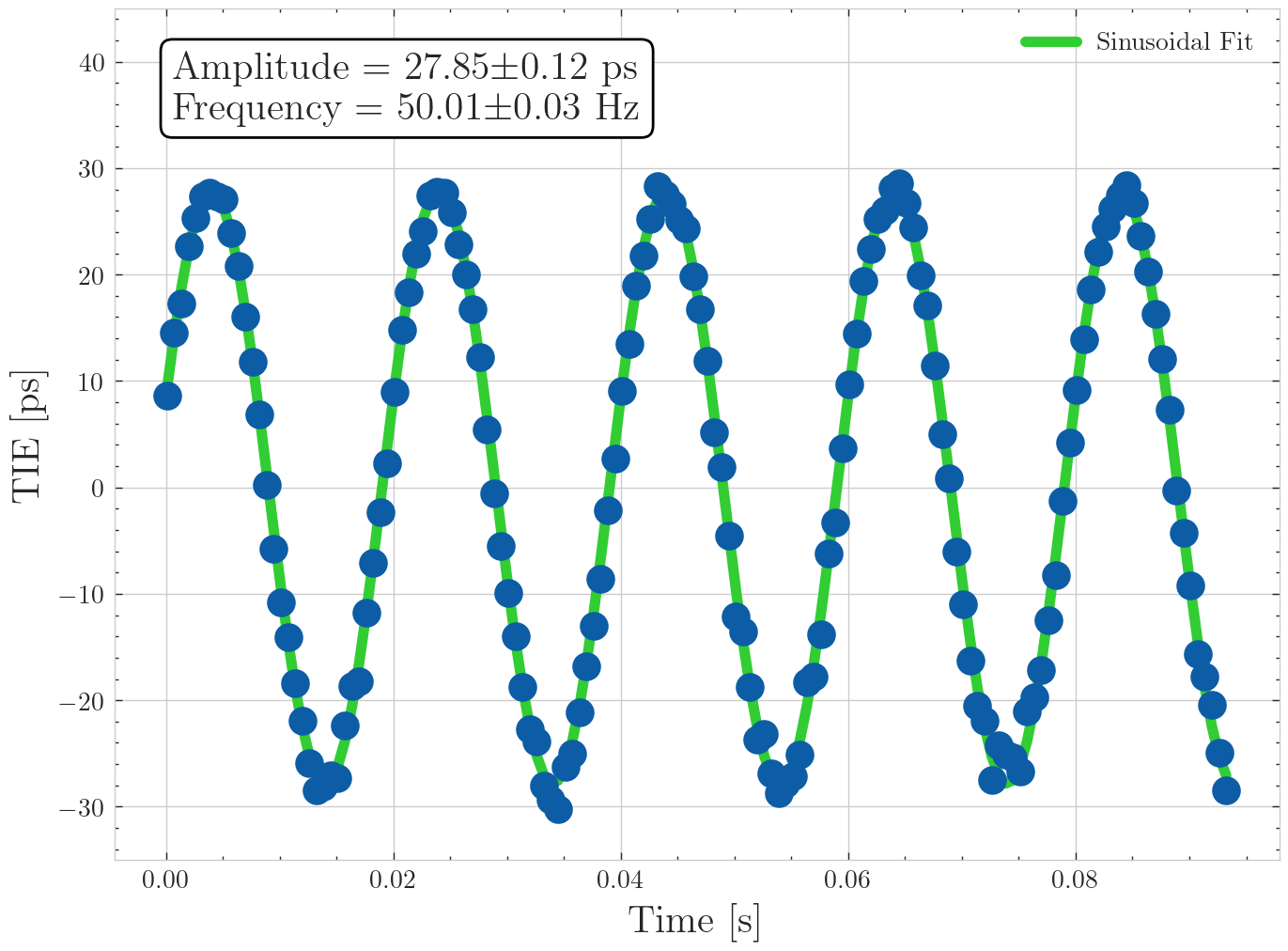}
    \end{minipage}
    \begin{minipage}[t]{0.48\linewidth}
        \centering
        \includegraphics[width=0.95\linewidth]{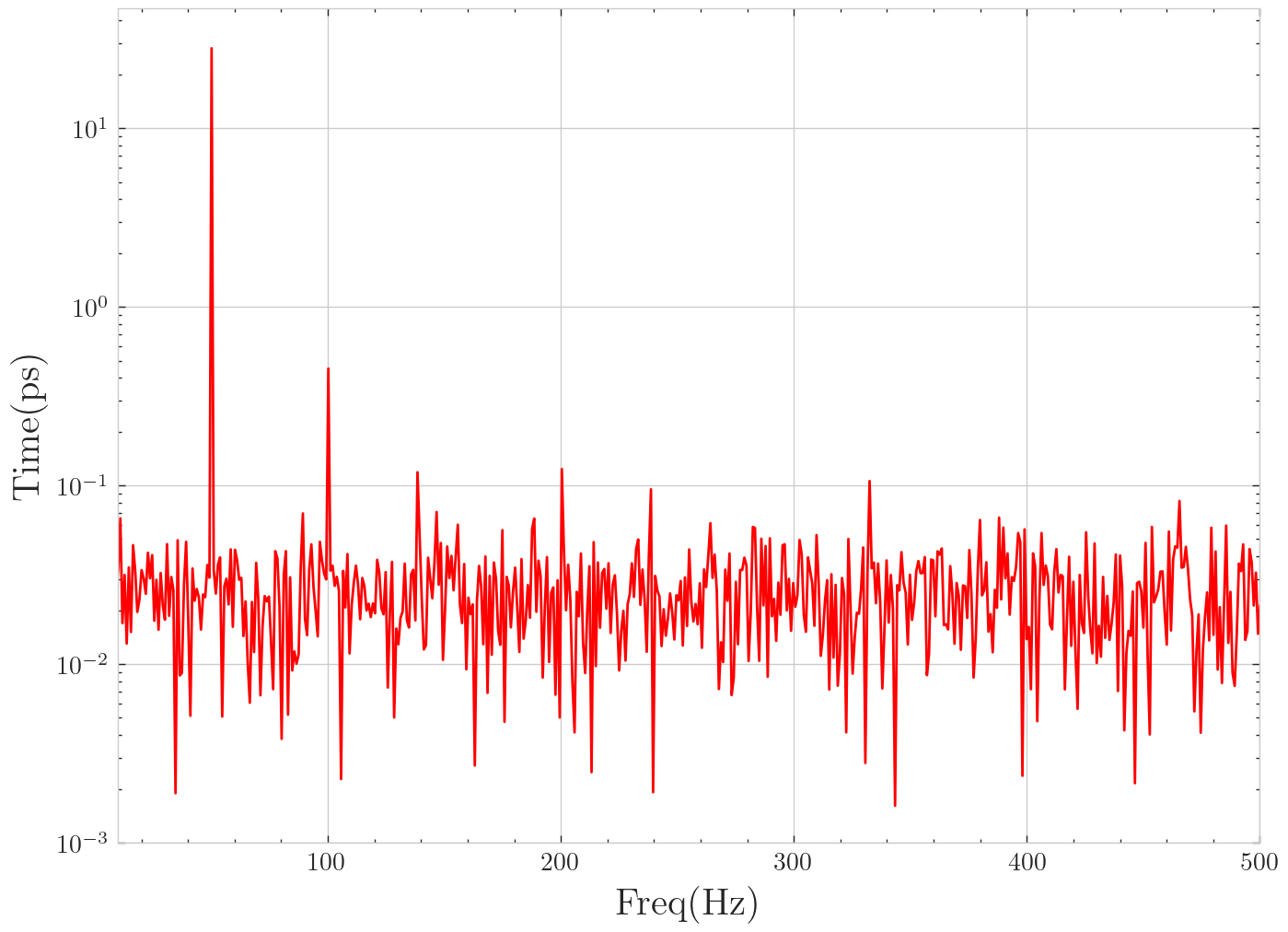}
    \end{minipage}
        \caption{Left: Time Interval Error recovered when a harmonic 50 Hz jitter with an amplitude of 28~ps is injected onto a 160 MHz digital clock. Right: FFT of the TIE signal recovered where the peak at 50 Hz and higher monotones can be seen.}
        \label{fig:TIEvsTime_100k}
  \end{figure}

\begin{figure}
    \centering
    \includegraphics[width=\textwidth]{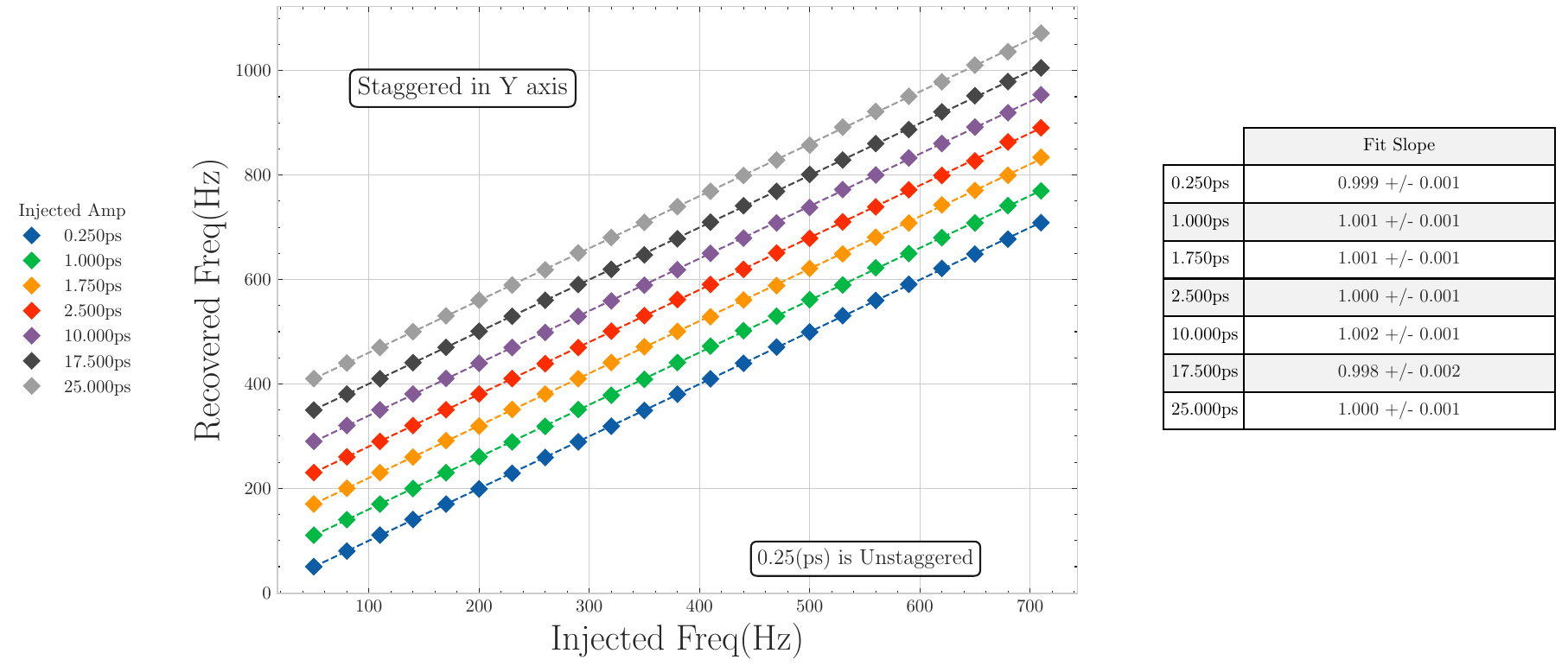}
    \caption{Recovered Noise Frequency vs Injected Noise Frequency 0.25 to 25~ps}
    \label{fig:freq_100k}
\end{figure}

For these tests the offset parameter (as defined in \cref{nu_ddmtd}) was set to N=100,000 corresponding to a $\nu_{max}$ of 1.6~kHz at a carrier clock frequency of 160 MHz.
\cref{fig:TIEvsTime_100k} shows the results from the DDMTD for a injected sinusoidal variation of 50~Hz at a noise amplitude of 25~ps. We observe the 50~Hz signal and its higher monotones in the FFT.
\cref{fig:freq_100k} shows the response of the system when the input jitter was modulated from 40 Hz to 750~Hz with amplitudes ranging from 0.25~ps to 25~ps.

\begin{figure}
    \centering
    \includegraphics[width=\textwidth]{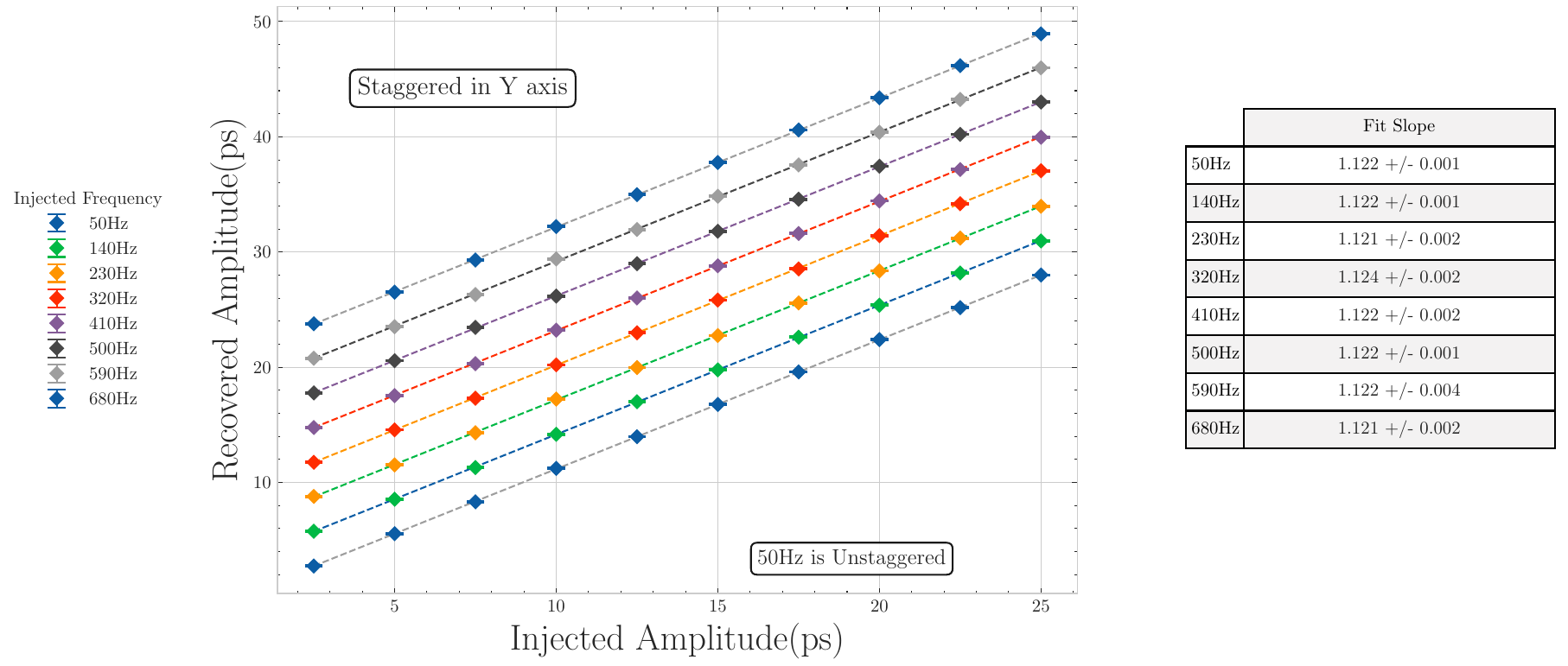}
    \caption{Recovered noise amplitudes for injected noise amplitudes between 2.5 to 25 ps; N=100k. The plots are staggered for readability.}
    \label{fig:Amp0.25_100k}
\end{figure}

\begin{figure}
    \centering
    \includegraphics[width=\textwidth]{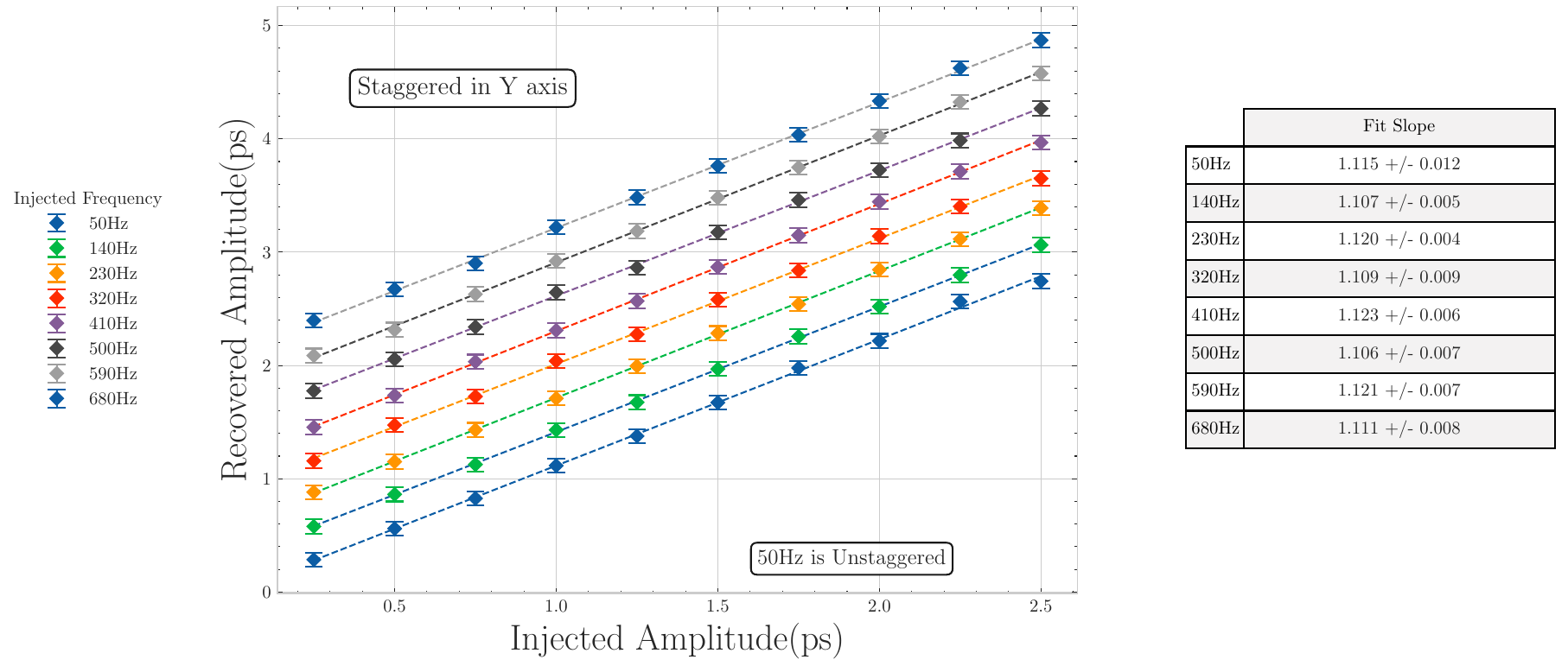}
    \caption{Recovered noise amplitudes for injected noise amplitudes between 0.25 to 2.5 ps; N=100k. The plots are staggered for readability.}
    \label{fig:Amp2.5_100k}
\end{figure}

Similar observations are shown in \cref{fig:Amp0.25_100k,fig:Amp2.5_100k}, where the recovered noise amplitudes were compared against the injected noise amplitudes. 

\begin{figure}
    \centering
    \includegraphics[width=\textwidth]{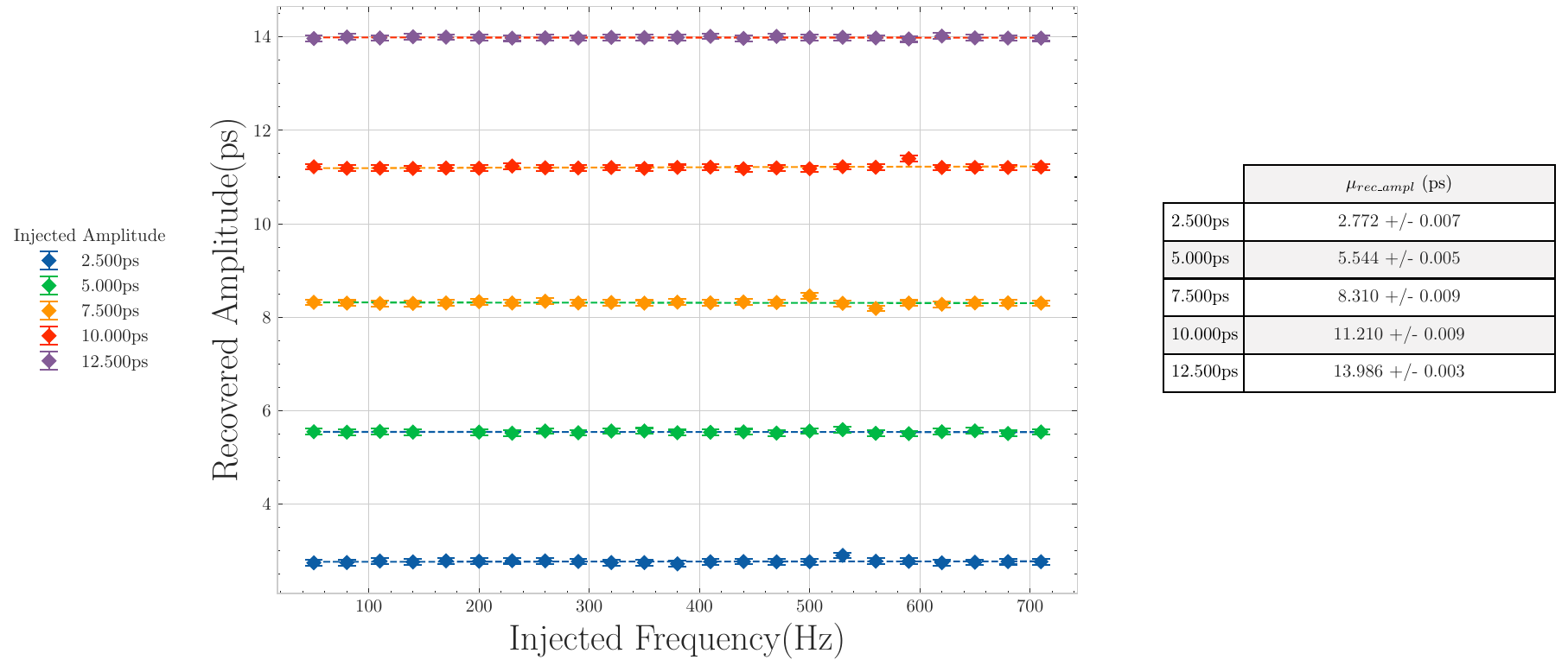}
    \caption{Stability of the recovered amplitude of the injected harmonic noise as a function of frequency, for injected amplitudes between 2.5 and 12.5 ps, measured with N=100k.}
    \label{fig:Stab0.25_100k}
\end{figure}

\begin{figure}
    \centering
    \includegraphics[width=\textwidth]{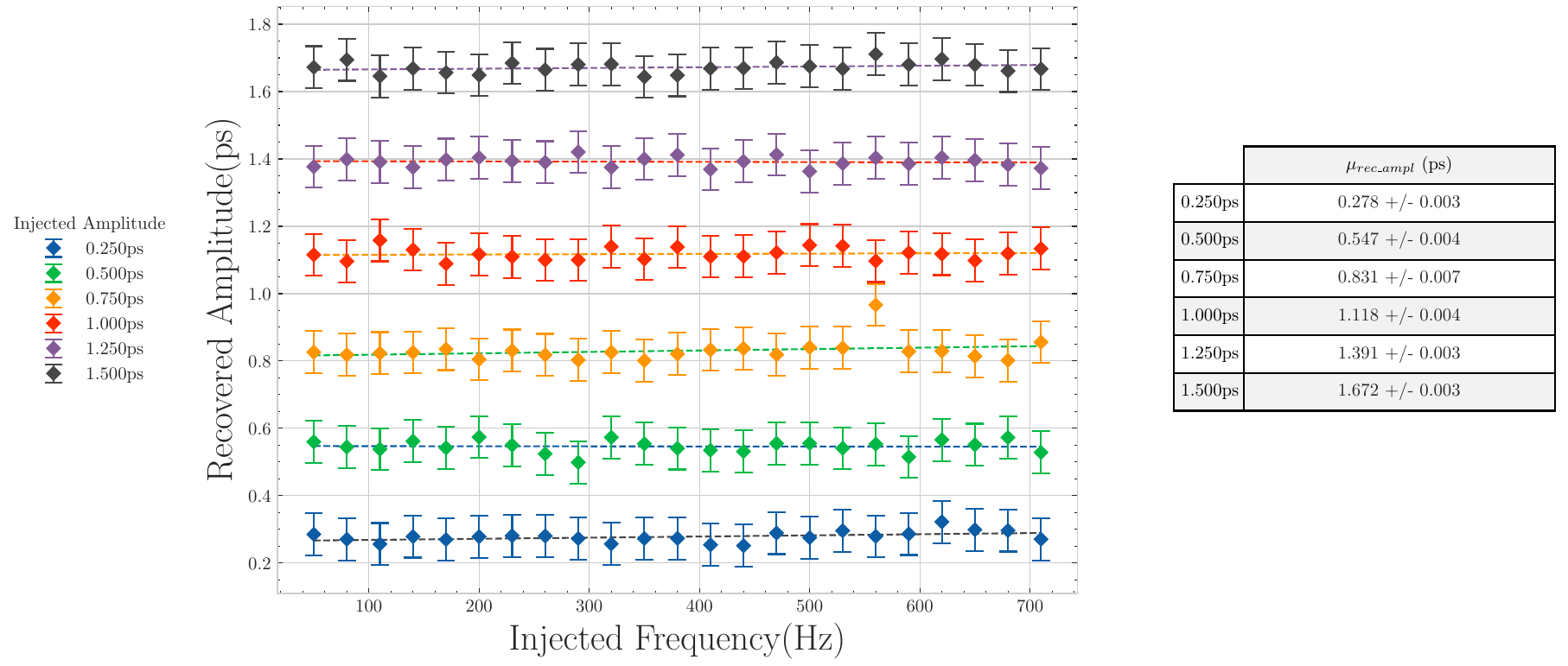}
    \caption{Stability of the recovered amplitude of the injected harmonic noise as a function of frequency, for injected amplitudes between 0.25 and 2.5 ps, measured with N=100k.}
    \label{fig:Stab2.5_100k}
\end{figure}

The response of the DDMTD system to different frequencies of injected noise are shown in \cref{fig:Stab0.25_100k,fig:Stab2.5_100k} and is observed to be flat across all the injected frequencies with a maximum standard deviation less than 100~fs. 


\section{Measurements made with the DDMTD}
\subsection{Effects of temperature on Front-End Electronics and Optical Fiber}

\begin{figure}
    \centering
    \includegraphics[width=\textwidth]{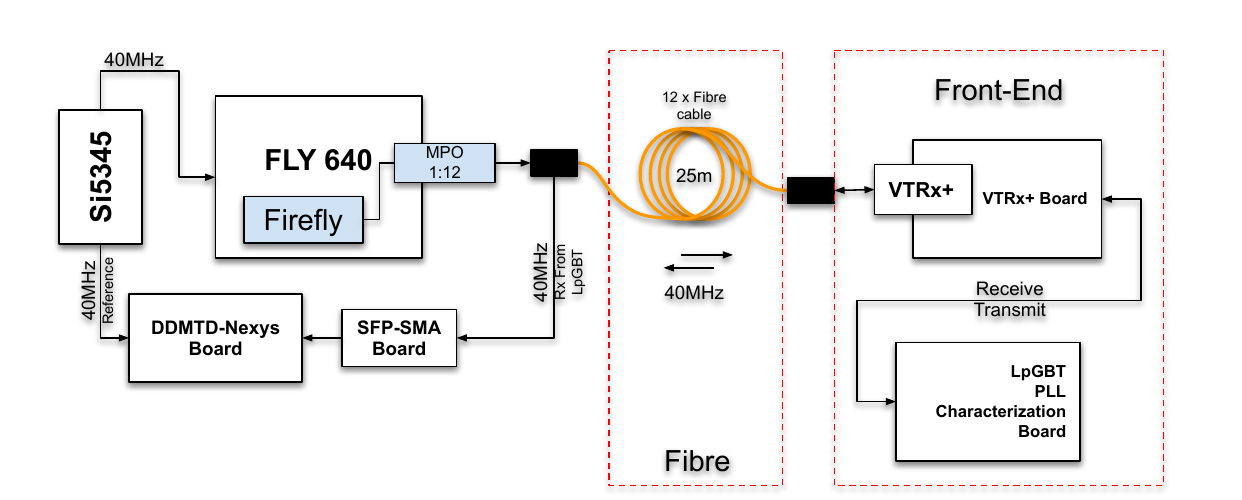}
    \caption{Configuration of the clock monitoring system used to measure the effect of temperature on an optical fiber and on the front-end emulator.}
    \label{fig:temp_schema}
\end{figure}

We used the clock-monitoring system to investigate temperature effects in transmission optical fibers and in a front-end emulator with an LpGBT and a VTRx+ optical transceiver \cite{VLB}.  
The configuration of the test system is shown in \cref{fig:temp_schema}. 

For these tests we used an Si5345 jitter attenuator as the source of a stable 40 MHz clock that was transmitted to a front end electronics emulator through a custom designed board, FLY 640 \cite{TWEPP}, on a 25~m long 12-channel multimode optical fiber. The front end emulator, consisting of a VTRX+ and an LpGBT characterization board, returned the 40~MHz clock signal back through the optical fiber to the DDMTD clock-monitoring system. A copy of the 40~MHz clock was also sent from the Si5344 board directly to the DDMTD as the reference clock.

With this setup, we measured the temperature effects on a 25~m optical fibre and on the front-end emulator by separately placing them inside a climate chamber. The parts of the set up that were kept inside the chamber are shown in \cref{fig:temp_schema} with red dotted boxes.

Compared to our qualification tests discussed above, the carrier frequency was 40~MHz rather than 160~MHz. At this frequency, with N=100k, only jitter frequencies up to 400~Hz are detectable, which is nevertheless much faster than any effects expected from temperature changes. 

\begin{figure}
\begin{subfigure}{0.5\textwidth}
  \centering
\includegraphics[width=\linewidth]{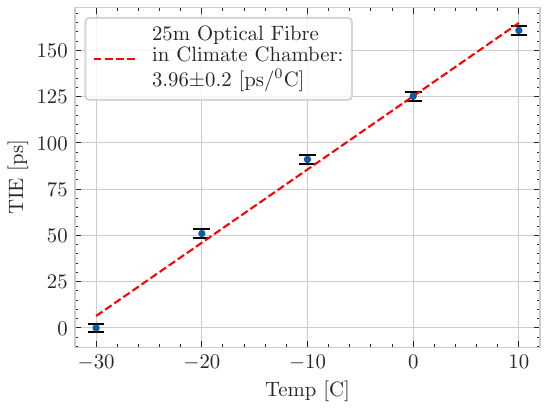}
  \caption{Optical Fiber in the climate chamber}
  \label{fig:fibre_temp}
\end{subfigure}
\hfill
\begin{subfigure}{0.5\textwidth}
  \centering
\includegraphics[width=\linewidth]{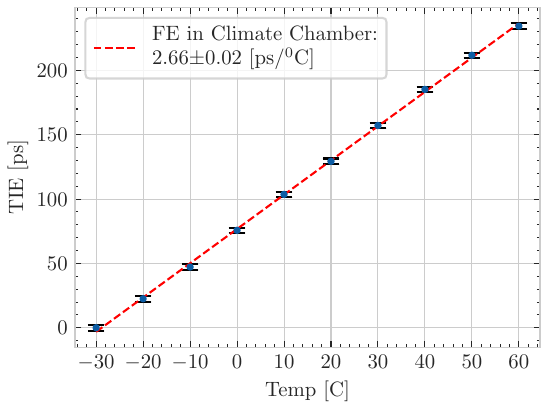}
  \caption{Front-End in the climate chamber}
  \label{fig:FE_temp}
\end{subfigure}
\caption{Time-interval error measured as a function of temperature for an optical fiber (left) and the front-end emulator (right).}
\label{fig:Temp_Measure}
\end{figure}

The first test was performed with the up-link and down-link fibres in the temperature-controlled chamber and the clock monitoring system was used to track changes in the signal delay time as the fiber temperature was changed.  The results of these tests are shown in \cref{fig:fibre_temp}. 
We observe a change of $\approx 4$ ps/$^{\circ}$C.
Assuming that the up-link and down-link shifts are symmetric, we measure the delay change of an optical clock signal propagating in multi-mode fiber to change by 0.08 $\pm$ 0.01~ps/m$\cdot^{\circ}$C.

In the second test we investigated the effect of temperature changes on the front-end emulator. \cref{fig:FE_temp} shows the time interval error as the temperature of the emulator board was changed from -30$^{\circ}$C to 60$^{\circ}$C in steps of 10$^{\circ}$C. 
Assuming a symmetric up-link and down-link delay, we observe a delay coefficient of $\approx$ 1.3~ps/$^{\circ}C$. 

These results are consistent with the measurements made with a digital oscilloscope and a phase noise analyser. 

\begin{center}
     \begin{figure}
    \includegraphics[width=\textwidth]{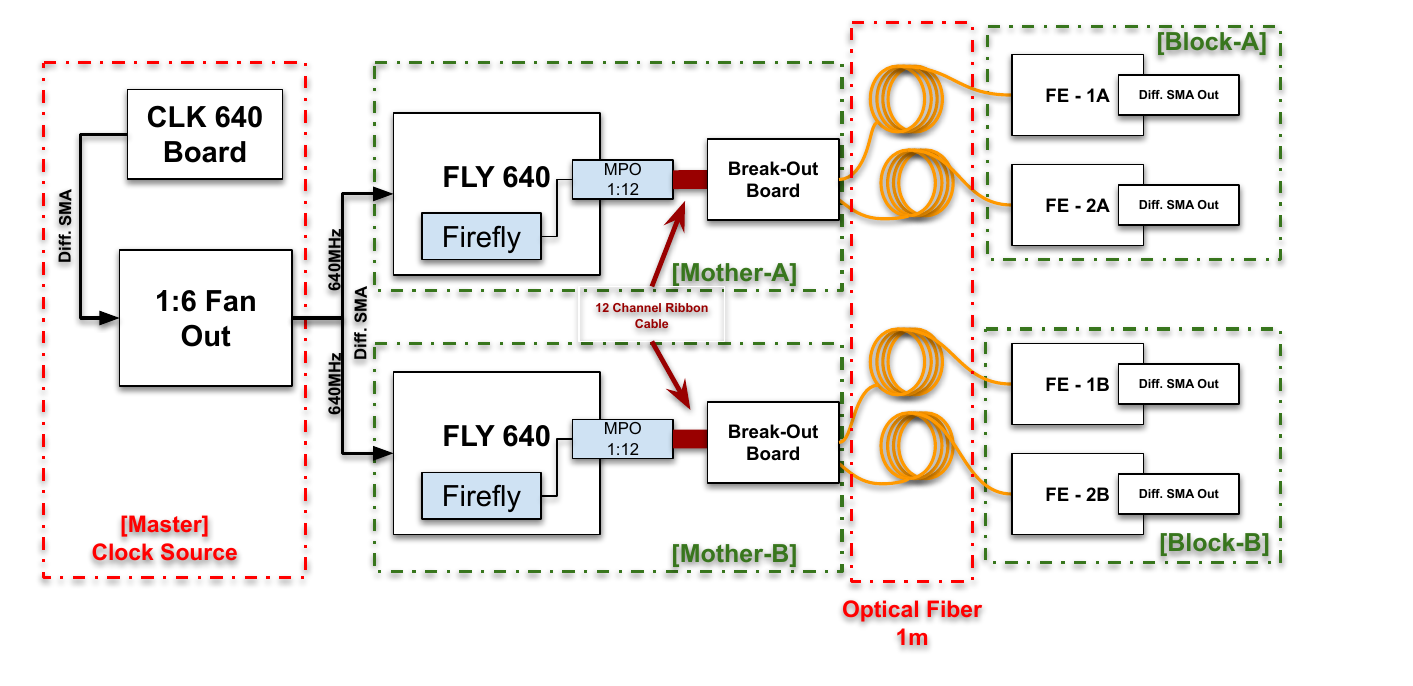}
    \caption{Schematic of the Pure Clock Distribution System.}
    \label{fig:pure_clock}
  \end{figure}
\end{center}
\subsection{Tests with a Pure Clock Distribution System}
As part of our investigation into the problem of distributing a precision clock we have designed and tested a scalable pure clock distribution system using high-performance, low-jitter, off-the-shelf components. Full details of the system may be found in~\cite{TWEPP}. In the system, a 640 MHz low-jitter digital clock is generated and distributed to two distribution boards, where copies of the clock are made using 1-6 fan-out and distributed using 12-channel SAMTEC Firefly Tx modules. The optical signal from the Fireflies is sent via 100~m long multi-mode fibers to front-end emulator boards (FE), where the optical signal is converted back to an electrical signal and divided by four and made available on SMC connectors. The system is shown schematically in \cref{fig:pure_clock}. 

We report here on tests made with this system using the DDMTD clock monitor with N=10k and 100k. 
The maximum frequency of the jitter that we are sensitive to is 16~kHz with N=10k, and 1.6~kHz with N=100k for a clock frequency of 160~MHz.
First we measured with the DDMTD the noise floor of the Si5344 in the "Master" by comparing two output clocks set at 160~MHz, where we obtained standard deviations of 0.3~ps and 0.4~ps for 100K and 10K, respectively. In both cases the noise level in the frequency domain was flat between 1 Hz and the upper limit of the measurement, 1.6~kHz for N = 100k and 16~kHz for N = 10k.

  \begin{figure}[hbt]
     \begin{subfigure}{0.49\textwidth}
    \includegraphics[width=0.95\linewidth]{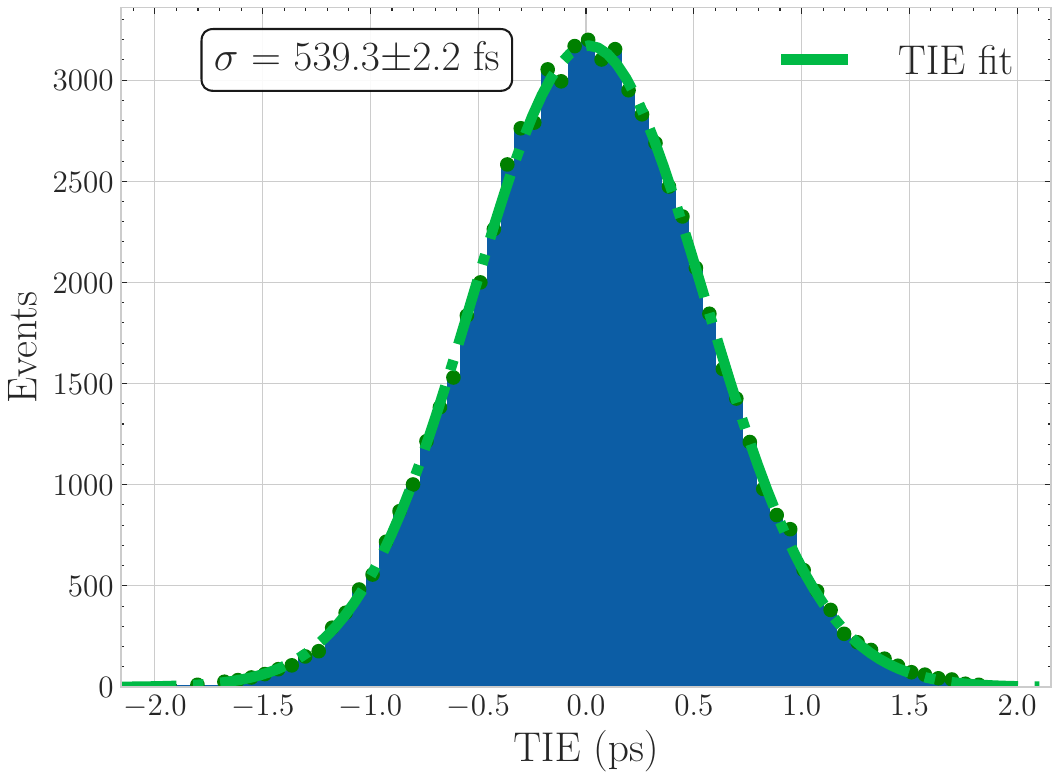}
    \caption{TIE :: $\sigma = 0.5$ ps}
    \label{fig:TIE_Cousins FE1-FE1}
  \end{subfigure}
  \hfill
    \begin{subfigure}{0.49\textwidth}
    \includegraphics[width=0.95\linewidth]{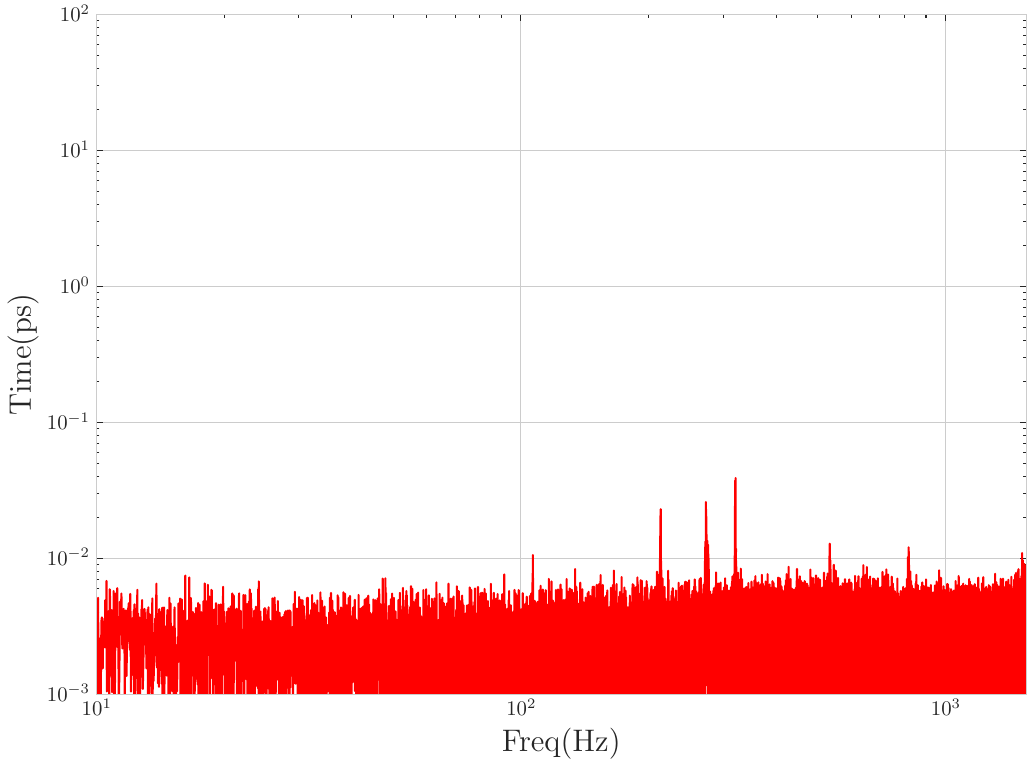}
    \caption{ FFT of the TIE}
    \label{fig:FFT_Cousins FE1-FE1}
  \end{subfigure}
  \caption{Measurement of the TIE at 160 MHz between the outputs FE-1A and FE-1B of the clock distribution system made with the DDMTD with N=100,000.}
  \label{fig:Cousins FE1-FE1}

  \begin{subfigure}{0.5\textwidth}
    \includegraphics[width=\linewidth]{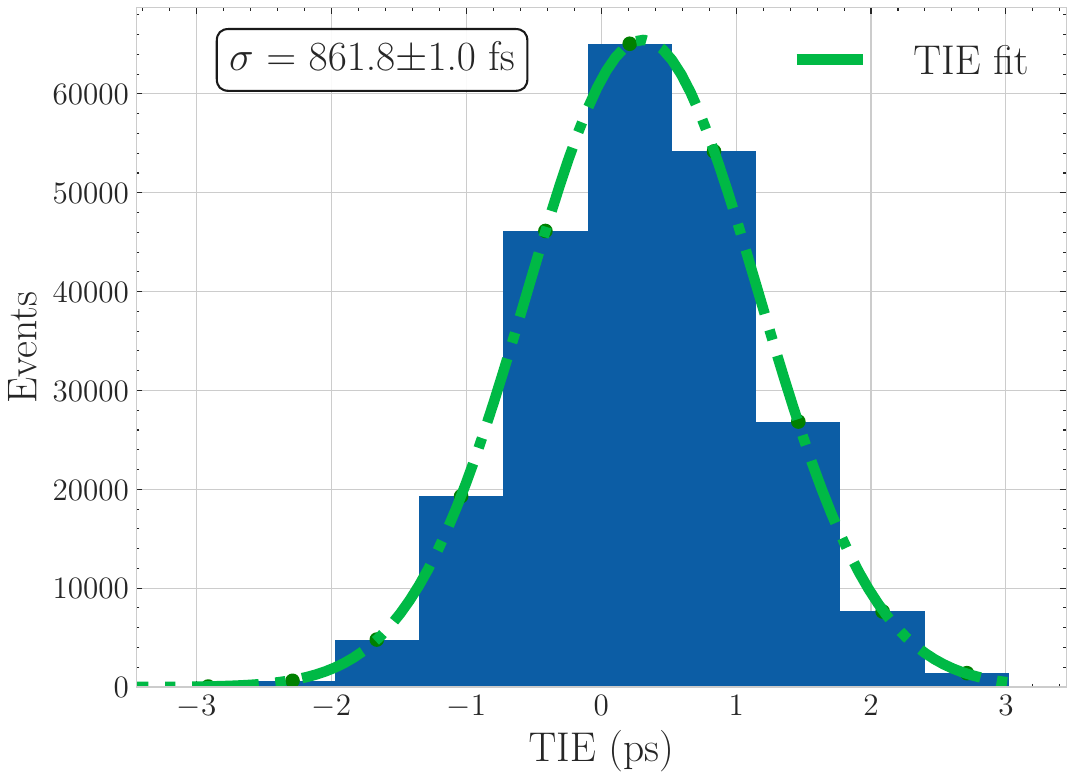}
    \caption{TIE :: $\sigma = 0.9$~ps}
    \label{fig:10kTIE_Cousins FE1-FE1}
  \end{subfigure}
  \hfill
    \begin{subfigure}{0.5\textwidth}
    \includegraphics[width=\linewidth]{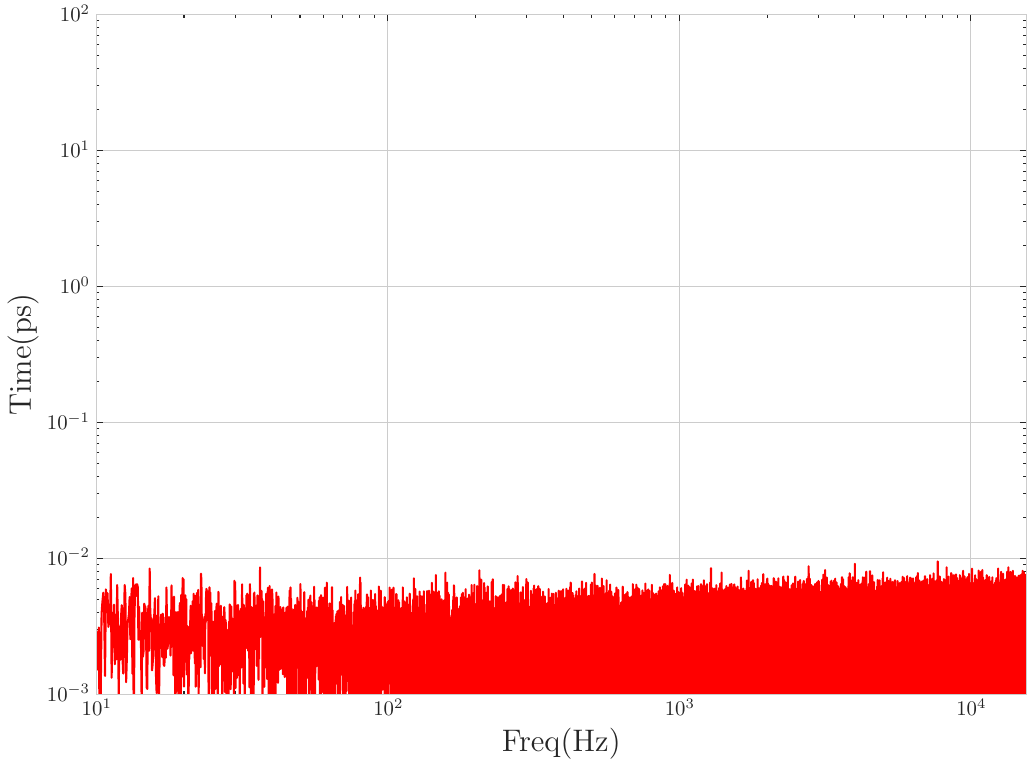}
    \caption{ FFT of the TIE}
    \label{fig:10kFFT_Cousins FE1-FE1}
  \end{subfigure}
  \caption{Measurement of the TIE at 160 MHz between the outputs FE-1A and FE-1B of the clock distribution system made with the DDMTD with N=10,000.}
  \label{fig:10kCousins FE1-FE1}
  \end{figure}

When we compared two output clocks from the same FE board, supplied by the same FLY 640 boards,
we obtained a standard deviation of 0.6~ps and 1.0~ps for N = 100k and 10k, respectively, and a flat response in frequency domain.
Comparing clocks distributed to two separate FE boards, which is comparable to how two reference clocks would be distributed in an experiment, we obtained the TIE distributions shown in \crefrange{fig:Cousins FE1-FE1}{fig:10kCousins FE1-FE1} for channel 1 in each FE board, with a standard deviations of 0.5~ps and 0.8~ps for N = 100K and 10K, respectively. Similar results were obtained comparing different combinations of the output clocks. The summary of the results obtained are given in \cref{table:Summary_PureClock}.

\begin{table}[htb]
\centering
\begin{tblr}{
  colspec = {c|c|c|c},
  cell{1}{3,5} = {c=2}{c}
}
\hline[2pt]
  \textit{\large $u_{1}$}  & \textit{\large $u_{2}$}   & $\sigma_{TIE}$(ps) &     \\
\hline[1pt]
       &  &   N=10k &  N=100k\\
\cline{3-5}  
Master-1 & Master-2 & 0.5   & 0.3\\
FE-1A    & FE-2A    & 1.1   & 0.7\\
FE-1B    & FE-2B    & 0.9   & 0.5\\
FE-1A    & FE-1B    & 0.9   & 0.5\\
FE-2A    & FE-2B    & 1.1   & 0.6\\
\hline[2pt]
\end{tblr}
\caption{Time-interval error measurements of the Pure Clock Distribution System using the DDMTD circuit.} \label{table:Summary_PureClock}
\end{table}

\section{Summary}
We describe the design and operation of a digital dual mixer time different (DDMTD) circuit built with discrete radio-frequency components. To characterise the circuit, harmonic noise with amplitudes ranging from 0.25~ps to 25~ps and frequencies ranging from 50~Hz to 710~Hz was injected onto a 160~MHz digital clock.   With the DDMTD we were able to recover the noise frequency of the injected noise with a precision of $0.2\%$ and
the measured amplitudes
of the injected noise were found to be stable across different injection frequencies, with a maximum standard deviation of less that 100~fs. 

 
Tests made with a pure clock distribution system had a maximum standard deviation of 1.1~ps between two parallel distributions systems with 100~m multi-mode fibers in each path.  We have also measured the delay of an optical clock signal propagating in a multi-mode fiber to be 0.08 $\pm$ 0.01~ps/m$\cdot^{\circ}$C.

 


\acknowledgments
We are grateful the US DOE Office of High Energy Physics for their support under the awards DE-SC0020185 and DE-SC0011845, and to the CERN High Precision Timing Laboratory for the use of their facilities. 
\bibliography{ref}
\bibliographystyle{JHEP}


\end{document}